\documentclass[letterpaper,twocolumn,10pt]{article}
\usepackage{usenix2019_v3}

\usepackage{booktabs} 
\usepackage{amssymb}
\usepackage{pifont} 
\usepackage{diagbox}
\usepackage{listings}
\usepackage{xcolor}
\usepackage{tabulary}
\usepackage{seqsplit}
\usepackage[draft,finalizecache,cachedir=.]{minted}

\usepackage{enumitem}
\usepackage{libertine}
\usepackage{graphicx}
\usepackage{multirow}
\usepackage{adjustbox}
\usepackage{url}
\usepackage{hyperref}
\usepackage[hyphenbreaks]{breakurl}
\usepackage[]{hyperref}         
\hypersetup{                    
  colorlinks,
  linkcolor={blue!70!black},
  citecolor={red!70!black},
  urlcolor={blue!70!black}
}

\newcommand{\code}[1]{{\fontfamily{cmtt}\fontseries{m}\fontshape{n}\selectfont\small{#1}}}

\newcommand{\oursystemtext}{Siren}
\newcommand{\oursystem}{\code{{\oursystemtext}}}
\newcommand*{\prompt}[1]{\textsf{\textbf{#1}}}
\newcommand{\tab}{\hspace*{1em}}

\usepackage[utf8]{inputenc}
\usepackage[T1]{fontenc}
\usepackage{authblk}

\makeatletter
\def\PYGbw@reset{\let\PYGbw@it=\relax \let\PYGbw@bf=\relax%
    \let\PYGbw@ul=\relax \let\PYGbw@tc=\relax%
    \let\PYGbw@bc=\relax \let\PYGbw@ff=\relax}
\def\PYGbw@tok#1{\csname PYGbw@tok@#1\endcsname}
\def\PYGbw@toks#1+{\ifx\relax#1\empty\else%
    \PYGbw@tok{#1}\expandafter\PYGbw@toks\fi}
\def\PYGbw@do#1{\PYGbw@bc{\PYGbw@tc{\PYGbw@ul{%
    \PYGbw@it{\PYGbw@bf{\PYGbw@ff{#1}}}}}}}
\def\PYGbw#1#2{\PYGbw@reset\PYGbw@toks#1+\relax+\PYGbw@do{#2}}

\expandafter\def\csname PYGbw@tok@gu\endcsname{\let\PYGbw@bf=\textbf}
\expandafter\def\csname PYGbw@tok@ch\endcsname{\let\PYGbw@it=\textit}
\expandafter\def\csname PYGbw@tok@gs\endcsname{\let\PYGbw@bf=\textbf}
\expandafter\def\csname PYGbw@tok@cm\endcsname{\let\PYGbw@it=\textit}
\expandafter\def\csname PYGbw@tok@gp\endcsname{\let\PYGbw@bf=\textbf}
\expandafter\def\csname PYGbw@tok@ge\endcsname{\let\PYGbw@it=\textit}
\expandafter\def\csname PYGbw@tok@cs\endcsname{\let\PYGbw@it=\textit}
\expandafter\def\csname PYGbw@tok@gh\endcsname{\let\PYGbw@bf=\textbf}
\expandafter\def\csname PYGbw@tok@ni\endcsname{\let\PYGbw@bf=\textbf}
\expandafter\def\csname PYGbw@tok@nn\endcsname{\let\PYGbw@bf=\textbf}
\expandafter\def\csname PYGbw@tok@s2\endcsname{\let\PYGbw@it=\textit}
\expandafter\def\csname PYGbw@tok@s1\endcsname{\let\PYGbw@it=\textit}
\expandafter\def\csname PYGbw@tok@nc\endcsname{\let\PYGbw@bf=\textbf}
\expandafter\def\csname PYGbw@tok@ne\endcsname{\let\PYGbw@bf=\textbf}
\expandafter\def\csname PYGbw@tok@si\endcsname{\let\PYGbw@bf=\textbf\let\PYGbw@it=\textit}
\expandafter\def\csname PYGbw@tok@nt\endcsname{\let\PYGbw@bf=\textbf}
\expandafter\def\csname PYGbw@tok@dl\endcsname{\let\PYGbw@it=\textit}
\expandafter\def\csname PYGbw@tok@sc\endcsname{\let\PYGbw@it=\textit}
\expandafter\def\csname PYGbw@tok@c1\endcsname{\let\PYGbw@it=\textit}
\expandafter\def\csname PYGbw@tok@kc\endcsname{\let\PYGbw@bf=\textbf}
\expandafter\def\csname PYGbw@tok@c\endcsname{\let\PYGbw@it=\textit}
\expandafter\def\csname PYGbw@tok@sx\endcsname{\let\PYGbw@it=\textit}
\expandafter\def\csname PYGbw@tok@err\endcsname{\def\PYGbw@bc##1{\setlength{\fboxsep}{0pt}\fcolorbox[rgb]{1.00,0.00,0.00}{1,1,1}{\strut ##1}}}
\expandafter\def\csname PYGbw@tok@kd\endcsname{\let\PYGbw@bf=\textbf}
\expandafter\def\csname PYGbw@tok@ss\endcsname{\let\PYGbw@it=\textit}
\expandafter\def\csname PYGbw@tok@sr\endcsname{\let\PYGbw@it=\textit}
\expandafter\def\csname PYGbw@tok@k\endcsname{\let\PYGbw@bf=\textbf}
\expandafter\def\csname PYGbw@tok@kn\endcsname{\let\PYGbw@bf=\textbf}
\expandafter\def\csname PYGbw@tok@cpf\endcsname{\let\PYGbw@it=\textit}
\expandafter\def\csname PYGbw@tok@kr\endcsname{\let\PYGbw@bf=\textbf}
\expandafter\def\csname PYGbw@tok@s\endcsname{\let\PYGbw@it=\textit}
\expandafter\def\csname PYGbw@tok@sh\endcsname{\let\PYGbw@it=\textit}
\expandafter\def\csname PYGbw@tok@ow\endcsname{\let\PYGbw@bf=\textbf}
\expandafter\def\csname PYGbw@tok@sb\endcsname{\let\PYGbw@it=\textit}
\expandafter\def\csname PYGbw@tok@sa\endcsname{\let\PYGbw@it=\textit}
\expandafter\def\csname PYGbw@tok@se\endcsname{\let\PYGbw@bf=\textbf\let\PYGbw@it=\textit}
\expandafter\def\csname PYGbw@tok@sd\endcsname{\let\PYGbw@it=\textit}

\makeatother

\makeatletter
\def\PYG@reset{\let\PYG@it=\relax \let\PYG@bf=\relax%
    \let\PYG@ul=\relax \let\PYG@tc=\relax%
    \let\PYG@bc=\relax \let\PYG@ff=\relax}
\def\PYG@tok#1{\csname PYG@tok@#1\endcsname}
\def\PYG@toks#1+{\ifx\relax#1\empty\else%
    \PYG@tok{#1}\expandafter\PYG@toks\fi}
\def\PYG@do#1{\PYG@bc{\PYG@tc{\PYG@ul{%
    \PYG@it{\PYG@bf{\PYG@ff{#1}}}}}}}
\def\PYG#1#2{\PYG@reset\PYG@toks#1+\relax+\PYG@do{#2}}

\expandafter\def\csname PYG@tok@gd\endcsname{\def\PYG@tc##1{\textcolor[rgb]{0.63,0.00,0.00}{##1}}}
\expandafter\def\csname PYG@tok@gu\endcsname{\let\PYG@bf=\textbf\def\PYG@tc##1{\textcolor[rgb]{0.50,0.00,0.50}{##1}}}
\expandafter\def\csname PYG@tok@gt\endcsname{\def\PYG@tc##1{\textcolor[rgb]{0.00,0.27,0.87}{##1}}}
\expandafter\def\csname PYG@tok@gs\endcsname{\let\PYG@bf=\textbf}
\expandafter\def\csname PYG@tok@gr\endcsname{\def\PYG@tc##1{\textcolor[rgb]{1.00,0.00,0.00}{##1}}}
\expandafter\def\csname PYG@tok@cm\endcsname{\let\PYG@it=\textit\def\PYG@tc##1{\textcolor[rgb]{0.25,0.50,0.50}{##1}}}
\expandafter\def\csname PYG@tok@vg\endcsname{\def\PYG@tc##1{\textcolor[rgb]{0.10,0.09,0.49}{##1}}}
\expandafter\def\csname PYG@tok@vi\endcsname{\def\PYG@tc##1{\textcolor[rgb]{0.10,0.09,0.49}{##1}}}
\expandafter\def\csname PYG@tok@vm\endcsname{\def\PYG@tc##1{\textcolor[rgb]{0.10,0.09,0.49}{##1}}}
\expandafter\def\csname PYG@tok@mh\endcsname{\def\PYG@tc##1{\textcolor[rgb]{0.40,0.40,0.40}{##1}}}
\expandafter\def\csname PYG@tok@cs\endcsname{\let\PYG@it=\textit\def\PYG@tc##1{\textcolor[rgb]{0.25,0.50,0.50}{##1}}}
\expandafter\def\csname PYG@tok@ge\endcsname{\let\PYG@it=\textit}
\expandafter\def\csname PYG@tok@vc\endcsname{\def\PYG@tc##1{\textcolor[rgb]{0.10,0.09,0.49}{##1}}}
\expandafter\def\csname PYG@tok@il\endcsname{\def\PYG@tc##1{\textcolor[rgb]{0.40,0.40,0.40}{##1}}}
\expandafter\def\csname PYG@tok@go\endcsname{\def\PYG@tc##1{\textcolor[rgb]{0.53,0.53,0.53}{##1}}}
\expandafter\def\csname PYG@tok@cp\endcsname{\def\PYG@tc##1{\textcolor[rgb]{0.74,0.48,0.00}{##1}}}
\expandafter\def\csname PYG@tok@gi\endcsname{\def\PYG@tc##1{\textcolor[rgb]{0.00,0.63,0.00}{##1}}}
\expandafter\def\csname PYG@tok@gh\endcsname{\let\PYG@bf=\textbf\def\PYG@tc##1{\textcolor[rgb]{0.00,0.00,0.50}{##1}}}
\expandafter\def\csname PYG@tok@ni\endcsname{\let\PYG@bf=\textbf\def\PYG@tc##1{\textcolor[rgb]{0.60,0.60,0.60}{##1}}}
\expandafter\def\csname PYG@tok@nl\endcsname{\def\PYG@tc##1{\textcolor[rgb]{0.63,0.63,0.00}{##1}}}
\expandafter\def\csname PYG@tok@nn\endcsname{\let\PYG@bf=\textbf\def\PYG@tc##1{\textcolor[rgb]{0.00,0.00,1.00}{##1}}}
\expandafter\def\csname PYG@tok@no\endcsname{\def\PYG@tc##1{\textcolor[rgb]{0.53,0.00,0.00}{##1}}}
\expandafter\def\csname PYG@tok@na\endcsname{\def\PYG@tc##1{\textcolor[rgb]{0.49,0.56,0.16}{##1}}}
\expandafter\def\csname PYG@tok@nb\endcsname{\def\PYG@tc##1{\textcolor[rgb]{0.00,0.50,0.00}{##1}}}
\expandafter\def\csname PYG@tok@nc\endcsname{\let\PYG@bf=\textbf\def\PYG@tc##1{\textcolor[rgb]{0.00,0.00,1.00}{##1}}}
\expandafter\def\csname PYG@tok@nd\endcsname{\def\PYG@tc##1{\textcolor[rgb]{0.67,0.13,1.00}{##1}}}
\expandafter\def\csname PYG@tok@ne\endcsname{\let\PYG@bf=\textbf\def\PYG@tc##1{\textcolor[rgb]{0.82,0.25,0.23}{##1}}}
\expandafter\def\csname PYG@tok@nf\endcsname{\def\PYG@tc##1{\textcolor[rgb]{0.00,0.00,1.00}{##1}}}
\expandafter\def\csname PYG@tok@si\endcsname{\let\PYG@bf=\textbf\def\PYG@tc##1{\textcolor[rgb]{0.73,0.40,0.53}{##1}}}
\expandafter\def\csname PYG@tok@s2\endcsname{\def\PYG@tc##1{\textcolor[rgb]{0.73,0.13,0.13}{##1}}}
\expandafter\def\csname PYG@tok@nt\endcsname{\let\PYG@bf=\textbf\def\PYG@tc##1{\textcolor[rgb]{0.00,0.50,0.00}{##1}}}
\expandafter\def\csname PYG@tok@nv\endcsname{\def\PYG@tc##1{\textcolor[rgb]{0.10,0.09,0.49}{##1}}}
\expandafter\def\csname PYG@tok@s1\endcsname{\def\PYG@tc##1{\textcolor[rgb]{0.73,0.13,0.13}{##1}}}
\expandafter\def\csname PYG@tok@dl\endcsname{\def\PYG@tc##1{\textcolor[rgb]{0.73,0.13,0.13}{##1}}}
\expandafter\def\csname PYG@tok@ch\endcsname{\let\PYG@it=\textit\def\PYG@tc##1{\textcolor[rgb]{0.25,0.50,0.50}{##1}}}
\expandafter\def\csname PYG@tok@m\endcsname{\def\PYG@tc##1{\textcolor[rgb]{0.40,0.40,0.40}{##1}}}
\expandafter\def\csname PYG@tok@gp\endcsname{\let\PYG@bf=\textbf\def\PYG@tc##1{\textcolor[rgb]{0.00,0.00,0.50}{##1}}}
\expandafter\def\csname PYG@tok@sh\endcsname{\def\PYG@tc##1{\textcolor[rgb]{0.73,0.13,0.13}{##1}}}
\expandafter\def\csname PYG@tok@ow\endcsname{\let\PYG@bf=\textbf\def\PYG@tc##1{\textcolor[rgb]{0.67,0.13,1.00}{##1}}}
\expandafter\def\csname PYG@tok@sx\endcsname{\def\PYG@tc##1{\textcolor[rgb]{0.00,0.50,0.00}{##1}}}
\expandafter\def\csname PYG@tok@bp\endcsname{\def\PYG@tc##1{\textcolor[rgb]{0.00,0.50,0.00}{##1}}}
\expandafter\def\csname PYG@tok@c1\endcsname{\let\PYG@it=\textit\def\PYG@tc##1{\textcolor[rgb]{0.25,0.50,0.50}{##1}}}
\expandafter\def\csname PYG@tok@fm\endcsname{\def\PYG@tc##1{\textcolor[rgb]{0.00,0.00,1.00}{##1}}}
\expandafter\def\csname PYG@tok@o\endcsname{\def\PYG@tc##1{\textcolor[rgb]{0.40,0.40,0.40}{##1}}}
\expandafter\def\csname PYG@tok@kc\endcsname{\let\PYG@bf=\textbf\def\PYG@tc##1{\textcolor[rgb]{0.00,0.50,0.00}{##1}}}
\expandafter\def\csname PYG@tok@c\endcsname{\let\PYG@it=\textit\def\PYG@tc##1{\textcolor[rgb]{0.25,0.50,0.50}{##1}}}
\expandafter\def\csname PYG@tok@mf\endcsname{\def\PYG@tc##1{\textcolor[rgb]{0.40,0.40,0.40}{##1}}}
\expandafter\def\csname PYG@tok@err\endcsname{\def\PYG@bc##1{\setlength{\fboxsep}{0pt}\fcolorbox[rgb]{1.00,0.00,0.00}{1,1,1}{\strut ##1}}}
\expandafter\def\csname PYG@tok@mb\endcsname{\def\PYG@tc##1{\textcolor[rgb]{0.40,0.40,0.40}{##1}}}
\expandafter\def\csname PYG@tok@ss\endcsname{\def\PYG@tc##1{\textcolor[rgb]{0.10,0.09,0.49}{##1}}}
\expandafter\def\csname PYG@tok@sr\endcsname{\def\PYG@tc##1{\textcolor[rgb]{0.73,0.40,0.53}{##1}}}
\expandafter\def\csname PYG@tok@mo\endcsname{\def\PYG@tc##1{\textcolor[rgb]{0.40,0.40,0.40}{##1}}}
\expandafter\def\csname PYG@tok@kd\endcsname{\let\PYG@bf=\textbf\def\PYG@tc##1{\textcolor[rgb]{0.00,0.50,0.00}{##1}}}
\expandafter\def\csname PYG@tok@mi\endcsname{\def\PYG@tc##1{\textcolor[rgb]{0.40,0.40,0.40}{##1}}}
\expandafter\def\csname PYG@tok@kn\endcsname{\let\PYG@bf=\textbf\def\PYG@tc##1{\textcolor[rgb]{0.00,0.50,0.00}{##1}}}
\expandafter\def\csname PYG@tok@cpf\endcsname{\let\PYG@it=\textit\def\PYG@tc##1{\textcolor[rgb]{0.25,0.50,0.50}{##1}}}
\expandafter\def\csname PYG@tok@kr\endcsname{\let\PYG@bf=\textbf\def\PYG@tc##1{\textcolor[rgb]{0.00,0.50,0.00}{##1}}}
\expandafter\def\csname PYG@tok@s\endcsname{\def\PYG@tc##1{\textcolor[rgb]{0.73,0.13,0.13}{##1}}}
\expandafter\def\csname PYG@tok@kp\endcsname{\def\PYG@tc##1{\textcolor[rgb]{0.00,0.50,0.00}{##1}}}
\expandafter\def\csname PYG@tok@w\endcsname{\def\PYG@tc##1{\textcolor[rgb]{0.73,0.73,0.73}{##1}}}
\expandafter\def\csname PYG@tok@kt\endcsname{\def\PYG@tc##1{\textcolor[rgb]{0.69,0.00,0.25}{##1}}}
\expandafter\def\csname PYG@tok@sc\endcsname{\def\PYG@tc##1{\textcolor[rgb]{0.73,0.13,0.13}{##1}}}
\expandafter\def\csname PYG@tok@sb\endcsname{\def\PYG@tc##1{\textcolor[rgb]{0.73,0.13,0.13}{##1}}}
\expandafter\def\csname PYG@tok@sa\endcsname{\def\PYG@tc##1{\textcolor[rgb]{0.73,0.13,0.13}{##1}}}
\expandafter\def\csname PYG@tok@k\endcsname{\let\PYG@bf=\textbf\def\PYG@tc##1{\textcolor[rgb]{0.00,0.50,0.00}{##1}}}
\expandafter\def\csname PYG@tok@se\endcsname{\let\PYG@bf=\textbf\def\PYG@tc##1{\textcolor[rgb]{0.73,0.40,0.13}{##1}}}
\expandafter\def\csname PYG@tok@sd\endcsname{\let\PYG@it=\textit\def\PYG@tc##1{\textcolor[rgb]{0.73,0.13,0.13}{##1}}}

\makeatother


\PassOptionsToPackage{hyphens}{url}
\usepackage{hyperref}

\newcommand*{\affaddr}[1]{#1} 
\newcommand*{\affmark}[1][*]{\textsuperscript{#1}}
\newcommand*{\email}[1]{\normalsize{}{#1}}

\title{Towards a First Step to Understand the Cryptocurrency Stealing Attack on Ethereum}

\author{%
\normalsize{Zhen Cheng\affmark[1,2]$^\ast$, Xinrui Hou\affmark[2,4]\thanks{Co-first authors with equal contribution.}\hspace*{0.3em}, Runhuai Li\affmark[1,2], Yajin Zhou\affmark[1,2]\thanks{Corresponding author.}\hspace*{0.3em}, Xiapu Luo\affmark[3], Jinku Li\affmark[4] and Kui Ren\affmark[1,2]}\\
\normalsize{\affaddr{\affmark[1]Zhejiang University}}\\
\normalsize{\affaddr{\affmark[2]Alibaba-Zhejiang University Joint Research Institute of Frontier Technologies}}\\
\email{\textit{\{iimp,lirunhuai,yajin\_zhou,kuiren\}@zju.edu.cn}}\\
\normalsize{\affaddr{\affmark[3]The Hong Kong Polytechnic University}}\\
\email{\textit{csxluo@comp.polyu.edu.hk}}\\
\normalsize{\affaddr{\affmark[4]Xidian University}}\\
\email{\textit{xrhou@stu.xidian.edu.cn}},\hspace*{0.3em} \email{\textit{jkli@xidian.edu.cn}}
}

\begin{document}





\maketitle
\thispagestyle{plain}
\pagestyle{plain}
\begin{abstract}

We performed the first systematic study of
a new attack on Ethereum that steals
cryptocurrencies. The attack is due to the unprotected JSON-RPC endpoints
existed in Ethereum nodes that could be exploited by attackers to
transfer the Ether and ERC20 tokens to attackers-controlled accounts.
This study aims to shed light on the attack,
including malicious behaviors and profits of attackers. Specifically, we first designed and implemented a honeypot that could capture \textit{real} attacks in the wild. We then deployed the honeypot and reported results of the collected data in a period of six months. In total, our system captured more than $308$ million requests from $1,072$ distinct IP addresses. We further grouped attackers into $36$ groups with $59$ distinct Ethereum accounts. Among them, attackers of $34$ groups were stealing the Ether, while other $2$ groups were targeting ERC20 tokens. The further behavior analysis showed that attackers were following a three-steps pattern to steal the Ether. Moreover, we observed an interesting type of transaction called zero gas transaction, which has been leveraged by attackers to steal ERC20 tokens. At last, we estimated the overall profits of attackers. To engage the whole community, the dataset of captured attacks is released on https://github.com/zjuicsr/eth-honey.

\end{abstract}

\section{Introduction}
\label{sec:introduction}

The Ethereum network~\cite{yellowpaper} has attracted lots of attentions.
Users leverage this platform to transfer the Ether, the
official cryptocurrency of the network, or build DApps (decentralized 
applications) using smart contracts.
This, in turn, stimulates the popularity of the Ethereum network.

However, this popularity also attracts another type of users, i.e., attackers. They exploit the insecure setting of
Ethereum clients, e.g., Go-Ethereum~\cite{goethereum} and Parity~\cite{parity}, to steal cryptocurrencies. These clients, if \textit{not properly configured}, will expose a JSON-RPC endpoint \textit{without any authentication mechanism
enforced}. As a result,
they could be remotely reached by attackers to invoke many privileged methods to manipulate the Ethereum account, \textit{on behalf of the account holder who is using the client}~\footnote{Note that, the RPC interface is intended to be used with proper authentication.}. Though we have seen spot reports of stealing the Ether by hackers~\cite{attack1,attack2}, there is no systematic study of such an attack. In order words, not enough insights were provided on the attack.

To this end, we performed a systematic study to understand the cryptocurrency stealing attack on Ethereum. The purpose of our study is to shed lights on this attack, including detailed malicious behaviors, attacking strategies, and attackers' profits.
Our study is based on \textit{real} attacks in the wild captured by our system.


Specifically, we designed and implemented a system called {\oursystem}.
It consists of a honeypot that listens the default JSON-RPC port, i.e., \code{8545}, and accepts any incoming requests. To make our honeypot a reachable and valuable target, we register it as an Ethereum full node on the Internet and prepare a real Ethereum account that has  Ethers inside.
In order to implement an interactive honeypot, we use a real Ethereum client (Go-Ethereum in our work) as the back-end.
We then redirect all the incoming RPC requests to the back-end, except for those that may cause damages to our honeypot.
Our honeypot logs the information of the request, including the method that the attacker intends to invoke, and  its
parameters. For instance, our honeypot logs account addresses that attackers intend to transfer the stolen Ether into. We call these accounts as \textbf{malicious accounts}. We further crawl transactions from the Ethereum network and analyze them to identify \textbf{suspicious accounts}, which accept Ethers from malicious accounts. Though we are unaware of real owners of these accounts, they are most likely to be related to attackers since transactions exist from malicious accounts to them. We then estimate attackers' profits by calculating the income of malicious and suspicious accounts.

\smallskip \noindent
{\prompt{Findings}}\tab
We performed a detailed analysis on the data collected in a six-month period~\footnote{The dataset is released on https://github.com/zjuicsr/eth-honey.}. Some findings are in the following. 

\begin{itemize}
    \item \textbf{Attacks captured}\tab During a six-month period, our system captured $308.66$ million RPC requests from $1,072$ distinct
    IP addresses. Among these IP addresses, $9$ of them are considered as the main source of attacks, since they count around $83.8\%$ of all requests. One particular IP address 89.144.25.28 sent the most RPC requests, with a record of $101.73$ million requests in total. In order to hide their
    real IP addresses, attackers were leveraging the Tor network~\cite{torlist} to launch attacks. We also observed that some IP addresses of worldwide universities were probing our honeypot, though none of them were invoking methods to steal cryptocurrencies. Most of these IP addresses are from the PlanetLab nodes~\cite{PlanetLab}. 
    \item \textbf{Cryptocurrencies targeted}\tab We grouped attackers based on IP addresses and target Ethereum accounts.
    These accounts are the ones that attackers transferred the stolen cryptocurrencies into. In total, attackers are grouped into $36$ clusters with $59$ distinct malicious accounts. Among them, attackers of $34$ groups were stealing the Ether, while other $2$ groups were targeting
    ERC20 tokens.
    \item \textbf{Steal the Ether}\tab We observed that attackers were following a three-steps pattern to steal the Ether.
    They first probed potential victims,
    and then collected necessary information to construct parameters. After that, they launch the attack, either passively waiting for the account being unlocked by continuously polling the account state, or actively lunching a brute-force attack to crack the user's password to unlock the account. 
    \item \textbf{Steal ERC20 tokens}\tab Besides the Ether, attackers were also targeting ERC20 tokens. We observed a type of transaction called zero gas transaction, in which the sender of a transaction does not need to pay any transaction fee. We find that attackers were leveraging this type of transactions to steal tokens from \textit{fisher} accounts that intended to scam other users' Ethers, and exploiting the AirDrop mechanism to gain numerous bonus tokens. 
\end{itemize}


\smallskip \noindent
{\prompt{Contributions}}\tab
In summary, this paper makes the following main contributions:

\begin{itemize}[leftmargin=*]
    \item We designed and implemented a system that can capture \textit{real} attacks to steal cryptocurrencies through unprotected JSON-RPC ports of vulnerable Ethereum nodes.
    \item We deployed our system and reported attacks observed in a period of
    six months. 
    \item We reported various findings based on the analysis of collected
    data. The dataset is released to the community for further study.
\end{itemize}

The rest of the paper is structured as the follows: we introduce the background information in Section~\ref{sec:background} and present the methodology of our system to capture attacks in Section~\ref{sec:methodology}. We then analyze the attack in Section~\ref{sec:attackanalysis} and estimate profits in Section~\ref{sec:profit}, respectively. We discuss the limitation of our work in Section~\ref{sec:discussion} and related work in Section~\ref{sec:related}. We conclude our work in Section~\ref{sec:conclusion}.

\section{Background}
\label{sec:background}

In this section, we will briefly introduce the necessary background
about the Ethereum network~\cite{yellowpaper} to facilitate the understanding of this work.

\subsection{Ethereum Clients and the JSON-RPC}

\begin{sloppypar}

An Ethereum node usually runs a client software. There exist
several clients, e.g., Go-Ethereum and Parity. Both clients support remote procedure call (RPC) through the standard JSON-RPC API~\cite{jsonrpc}. When these clients are being started with a special flag, they will listen a specific port (e.g., \code{8545}), and accept RPC requests \textit{from
any host} \textit{without any authentication.}
After that, functions could be remotely invoked on behalf of the account holder of the client, including privileged ones to send transactions (or transfer cryptocurrencies). 
Note that, though the Ethereum network is a P2P network,
attackers can discover and reach vulnerable Ethereum nodes
directly through the HTTP protocol.

\subsection{Ethereum Accounts}
On the Ethereum platform, there exist two different types of accounts.
One is externally owned account (EOA), and another one is smart
contract account. An EOA account can transfer the Ether, the official currency in Ethereum, to another account. An EOA account can deploy a smart contract, which in turn creates another type of account, i.e., the smart contract account. A smart contract is a program that executes exactly as it is set up to by its creator (the smart contract developer). In Ethereum, the smart contract is usually programmed using the Solidity language~\cite{solidity}, and executes on a virtual machine called Ethereum Virtual Machine (EVM)~\cite{yellowpaper}. Both types of
accounts are denoted in a hexadecimal format. For instance, the account address \code{0x6ef57be1168628a2bd6c5788322a41265084408a} denotes an (attacker's) EOA account, while the address \code{0x87c9ea70f72ad55a12bc6155a30e047cf2acd798} denotes a smart contract.

ERC20 tokens~\cite{erc20tokenstandard} are digital tokens designed and used on the Ethereum platform, which could be shared, exchanged for other tokens or real currencies, e.g., US dollars. The community has created standards for issuing a new ERC20 token using the smart contract. For instance, the smart contract should implement a function called \code{transfer()} to transfer the token from one account to another, and a \code{balanceOf()} function to query the balance of the token. The values of ERC20 tokens vary for different tokens at different times. For instance, the market capitalization of the Minereum token~\cite{minereum} was more than 7 million US dollars~\cite{minereumcap} in August, 2017, and is
around $40,000$ US dollars in March, 2019.

\subsection{Transactions}
\label{subsec:transaction}


Transactions can be used to transfer the Ether, or invoke functions of a smart contract. Specifically, the \textit{to} field of a transaction denotes the destination, i.e., an EOA account or a smart contract. For a transaction to send the Ether, fields including \textit{gas} and \textit{gasPrice} specify the gas limit and the gas price of the transaction. Listing~\ref{list:attacker-address} (Section~\ref{sec:attackanalysis} on page~\pageref{list:attacker-address}) shows a real transaction to send the Ether to \code{0x63710c26a9be484581dcac1aacdd95ef628923ab}, a malicious EOA account captured by our system. If the transaction is used to invoke a function of a smart contract, then the \textit{data} field specifies the name and parameters of the function to be invoked. Note that,
a function is identified by a function signature, i.e., the four bytes of the Keccak hash of the canonical expression of the function prototype, including the function name, the parameter types.
Listing~\ref{list:attacker-rawtransaction} and \ref{list:transfer-method} (Section~\ref{sec:attackanalysis} on page~\pageref{list:attacker-rawtransaction})  shows the \textit{data} field and the signature of the invoked function and its prototype.

Sending a transaction consumes \textit{gas}, which is the name of the unit that measures the work needs to be done. It is similar to the use of a liter of fuel consumed when driving a car. The actual cost of sending a transaction (transaction fee) is calculated as the product of the consumed gas and the current gas price. The gas price is similar to the cost of each liter of fuel that is paid for filling up a car. The smallest unit of the Ether is {Wei}. A {Gwei} consists of a billion Wei, while an Ether consists of a billion Gwei. The amount of gas consumed in a transaction is accumulated during instruction execution. Since the operation of transferring the Ether is a sequence of fixed instructions, thus the consumed gas is always $21,000$.

The transaction fee is paid by the sender to the miner, who is responsible for packing transactions into blocks and executing smart contract instructions. To earn a higher transaction fee, miners tend to pack the transaction with a higher gas price. Specifically, the sender of a transaction can specify the gas price in the field \textit{gasPrice} (Section~\ref{sec:attackanalysis} on page~\pageref{list:attacker-address}) to boost the chance of the transaction being packed. We have observed a trend of higher gas price in the transactions to steal Ether (Figure~\ref{fig:gasprice} on page~\pageref{fig:gasprice}).

There are multiple RPC methods that could be remotely invoked to send (or sign) a transaction on behalf of the account holder, including \code{eth\_sendTransaction} and \code{eth\_signTransaction}. Note that the account needs to be unlocked before sending a transaction. This involves the process to enter the user's password. Otherwise, the invocation of these methods will fail. In other words, in order to successfully
steal the Ether from the victim's account, his or her account should be in the state of being unlocked.
In Section~\ref{subsec:analysisethstealing},
we will show that attackers continuously monitor the victim's
account until it is unlocked by the user, or launch a
brute-force attack using a predefined dictionary of popular passwords.

\end{sloppypar}

\section{Methodology}
\label{sec:methodology}

\begin{figure}[t]
    \includegraphics[width=0.45\textwidth]{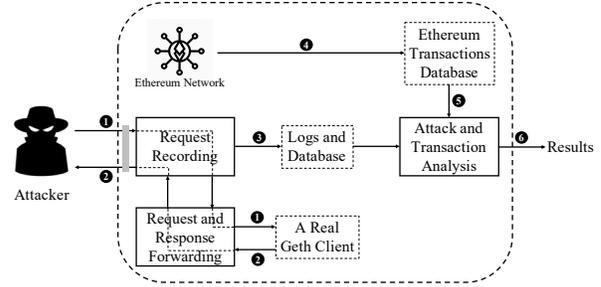}
    \caption{The overview of our system.}
    \label{fig:architecture}
\end{figure} 

Figure~\ref{fig:architecture} shows the architecture of our system.
In order to capture real attacks towards Ethereum nodes with unprotected HTTP JSON-RPC endpoints, we design and implement a system called {\oursystem}. Our system consists of a honeypot that listens the default JSON-RPC
port, i.e., \code{8545}. Any attempt to connect to this port of our honeypot
will be recorded. Our system forwards the request to the back-end Ethereum client(\ding{182}) and returns the results (\ding{183}).
The Ethereum client used in our system is the Go-Ethereum~\footnote{In this paper, if not otherwise specified, the Ethereum client we discussed is the Go-Ethereum.}. Our honeypot logs the RPC requests with parameters from the attacker, and then imports them into a database(\ding{184}). To further estimate profits of attackers, we crawl transaction records from the Ethereum network (\ding{185}), and identify suspicious accounts that are connected with attacker's accounts.
Our system combines the transaction records (\ding{186}) of both malicious and suspicious accounts to generate the final result (\ding{187}).

\subsection{Ethereum Honeypot}

In order to capture real attacks and understand attacking behaviors, we build a honeypot. It can interact with JSON-RPC requests that invoke APIs for malicious intents, e.g., transferring the Ether to attacker controlled malicious accounts.
Our honeypot logs the information of the API invocation, including the method name, parameters and etc., for later analysis. 
Moreover, our system confines the attacker's behaviors that APIs invoked cannot cause real damages to our honeypot.

To this end, we design a front-end/back-end architecture for our honeypot. Specifically, we implement a front-end that listens to the \code{8545} port and accepts any incoming HTTP JSON-RPC request from this port. We also have a real Ethereum client in the back-end, which runs as a full Ethereum node. This client
accepts any \textit{local} JSON-RPC request from the front-end. That means our real Ethereum client is not publicly available to the attackers. If the invoked API is inside a predefined whitelist, our front-end will forward the request to the back-end client, and then forward the response to the attacker. The APIs might bring a financial loss to our account are strictly forbidden. By doing so, our system protects the Ethereum node from being actually exploited, while at the same time, facilitates the information logging of the requests since all the requests need to go through the front-end.

However, there are several challenges that need to be addressed to make the system effective. For instance, the honeypot should behave like a real Ethereum node. Otherwise, attackers could be aware the existence of our honeypot
and do not perform malicious activities.
In the following, we will illustrate ways that our system leverages to attract attackers, and further describe how our honeypot works.

\smallskip \noindent
\prompt{Respond the probe requests}\tab The default port number of the HTTP JSON-RPC service of an Ethereum node is \code{8545}. Before launching a real attack, attackers usually send probe requests to check whether this port is actually open. For instance, the attacker invokes the \code{web3\_clientVersion} method to check whether it is a valid Ethereum node. The front-end of our honeypot accepts any incoming JSON-RPC request, and responds with valid results, by relaying responses from the back-end Ethereum client.

\smallskip \noindent
\prompt{Advise the existence of our Ethereum node}\tab In order to capture
attacks, our system needs to attract attackers. There are two  options for this purpose.
One option is that we passively wait for attackers by responding to the probing request. However, this strategy is not efficient since the chance that our honeypot is happened to be scanned is low, given the large space of valid IP addresses. 
The second option is to actively attract attackers. Specifically,
to make our Ethereum node (or our honeypot) visible to attackers, we register it on public websites that provide the list of full Ethereum nodes. The original purpose of this list is to speed up the discovery process of Ethereum
nodes in the P2P network. 
However, this list provides valuable information to attackers, since they can find potential targets without performing time- and resource-consuming port scanning process. It turns out this strategy is really effective. Our honeypot receives incoming probe requests shortly after being registered on the list.

\begin{sloppypar} 
\smallskip \noindent
\prompt{Pretend as a valuable target}\tab The main purpose of the attack is to steal cryptocurrencies. In order to make the attacker believe our honeypot is a valuable target, we create a real Ethereum account with the address \code{0xa33023b7c14638f3391d705c938ac506544b25c3} and transfer some amounts of Ether into this account. 
Since the Ethereum network is a public ledger, the amount of the Ether inside the account could be obtained by querying on the network. Our honeypot returns this account address to attackers if they invoke the \code{eth\_accounts} method to get a list of accounts owned by our Ethereum node. We also return the real amount of Ether inside this account to attackers if the \code{eth\_getBalance} method is invoked.
\end{sloppypar}

\smallskip \noindent
\prompt{Emulate a real transaction}\tab After obtaining the information of the account owned by our Ethereum node and the balance of the account, attackers tend to steal the Ether by transferring it to accounts they controlled (malicious accounts). For instance, they could invoke the \code{eth\_sendTransaction} method, which returns the hash value of a newly-created transaction. Attackers could check the return value of the method invocation to get the status of Ether transfer. To make the attacker believe that the transaction is being processing, while not actually transferring any Ether from our account, we do not actually execute the \code{eth\_sendTransaction} method. Instead, we log the parameters of this method invocation, and return a randomly generated hash value to the attacker.

\smallskip \noindent
\prompt{Log RPC requests}\tab Our honeypot logs the attacker's invoked methods, including the method name, parameters, along with the metadata of the attack, such as the IP address and the time. All the data will be saved into a log file, which will be imported into a database.

\subsection{Data Collection and Analysis}

After capturing attacks and malicious account addresses, we will estimate profits gained by attackers. Our system leverages transactions launched from these accounts to find more attacker-controlled accounts. For this purpose, we crawl the whole transactions from the Ethereum network.

Our system first downloads Ethereum transactions, then imports them into a database. After that, we can conveniently combine data captured by our honeypot and transactions from the Ethereum network to generate final analysis results.

\section{Attack Analysis}
\label{sec:attackanalysis}

In this section, we will illustrate the data we collected, the grouping process of attackers, and detailed information about attacks to steal the Ether and ERC20 tokens. 

\subsection{Data Overview}

\begin{figure}[t]
    \centering\includegraphics[width = 8cm]{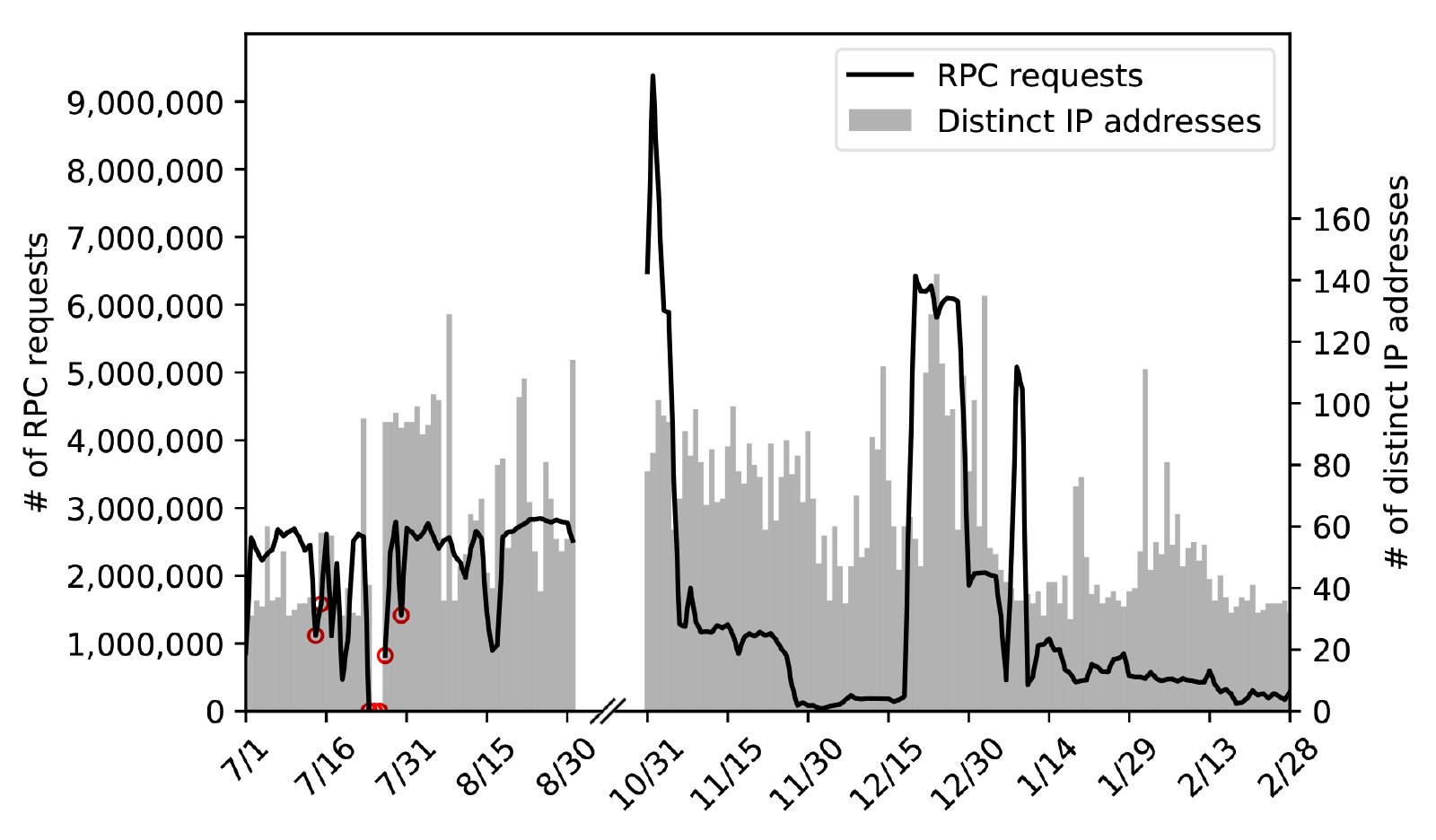}
    \caption{The number of daily RPC requests and distinct IP addresses captured by our honeypot. The seven red circles mean the data in these
    days is incomplete, either because our system was accidentally shut down, or the network was not stable.}   
    \label{fig:dailyamount}
\end{figure}

We deployed our system on a virtual machine in Alibaba cloud, and collected the data in a period of six months, i.e., from July 1 to August 31 in the year 2018, and November 1, 2018 to February 28, 2019. Unfortunately, there are three days (from July 24 to July 26) when the virtual machine was accidentally shut down, and four days (July 14, 15, 27 and 30) when the network was not stable. During these days, the data is either missing or incomplete. Figure~\ref{fig:dailyamount} shows the number of daily RPC requests and distinct IP addresses of these requests.

In total, our system observed $308.66$ million RPC requests from $1,072$ distinct IP addresses. In average, we received $1.72$ million RPC requests each day (excluding the incomplete data.) In terms of IP addresses, the average daily number is $62$. This number reached its peak value ($142$) on December 24, 2018.






Among these RPC requests, $9$ different IP addresses are main sources of attacks, given that they contribute most of RPC requests in our dataset. These $9$ IP addresses sent around $258.70$ million requests in total, which counts around $83.8\%$ of all requests. It's worth noting that, the IP address 89.144.25.28 sent the most RPC requests. It sent $101.73$ million requests in total, accounting for $33.0\%$ of the requests we received. We believe such an aggressive behavior is to increase the possibility of stealing the Ether, since the time window to transfer the Ether only exists when the user account is unlocked. We also observed that attacks from this IP address ceased in several days, e.g., from August 15 to 17. 

\smallskip \noindent
\prompt{RPC requests from universities worldwide}\tab Interestingly, some RPC requests are from IP addresses that belong to universities. Specifically, our honeypot received requests from $66$ IP addresses of $39$ universities in $13$ countries or regions. Among them, $37$ IP addresses are from universities in the USA. For instance, two IP addresses (146.57.249.98 and 146.57.249.99) belong to the University of Minnesota. We further used the reverse DNS lookup command to obtain the domain name associated with these IP addresses. It turns out that all of them are associated with the PlanetLab~\cite{PlanetLab}. For instance, the domain names of the previous two IP addresses are \code{planetlab1.dtc.umn.edu} and \code{planetlab2.dtc.umn.edu}, respectively.
Requests from these IP addresses are not performing malicious activities, e.g., transferring the Ether to other accounts. Most of them are merely probing our honeypot for information collection, e.g., invoking \code{eth\_getBlockByNumber}. Though the exact intention of collecting such information is unknown, we believe these requests are mainly for a research purpose. 

\smallskip \noindent
\prompt{Abuse of the Tor network and cloud services}\tab
Attackers are leveraging the Tor network and cloud services to hide their identities. For instance, some IP addresses belong to popular cloud services, e.g., Amazon, DigitalOcean and etc.
Among the $1,072$ distinct IP addresses, $370$ of them are identified as Tor gateways~\cite{torlist} ($299$ of them performed malicious behaviors, e.g., trying to steal the Ether.) All these IP addresses belong to the second group (Section~\ref{subsec:grouping}.) They are from $104$ different ISPs in $39$ countries. Using the Tor network to hide the real IP addresses make the tracing of attackers more difficult.

\begin{listing}[t]
    \usemintedstyle{bw}
    \begin{minted}[frame=single,
                    fontsize=\tiny,
                    tabsize=2,escapeinside=||]{js}
// Date: Jul 1 20:44:09 GMT+08:00 2018
// Source IP: 89.144.25.28
{
    "jsonrpc": "2.0", 
    "method": "eth_sendTransaction", 
    "params": [{
        //The account address of our honeypot.
        "from": "0xa33023b7c14638f3391d705c938ac506544b25c3", 
        //Attacker's account address.
        "to": "0x63710c26a9be484581dcac1aacdd95ef628923ab", 
        "gas": "0x5208", 
        "gasPrice": "0x199c82cc00", 
        "value": "0x2425f024b7fd000", 
    }], 
    "id": 739296
}
    \end{minted}
    \caption{The captured attack and the associated account address in the \textit{to} field.} 
    \label{list:attacker-address}
\end{listing}

\subsection{Grouping Attackers Accounts}
\label{subsec:grouping}

After collecting the data, our next step is to group attackers. However, 
due to the anonymity property of the Ethereum network, it is hard
to group them based on their identities.
In this paper, we take the following ways to group attackers.



\begin{sloppypar}

First, we \textit{directly} retrieve attackers' Ethereum accounts through the parameters of RPC requests,
and group them based on these accounts. For instance, the parameter \textit{to} of the method \code{eth\_sendTransaction} denotes the destination account of a transaction that the Ether will be transferred into. Attackers use this method to transfer (steal) the Ether to their controlled accounts. Listing~\ref{list:attacker-address} shows the parameters (in the JSON format) of a captured malicious transaction launched by an attacker. The value of the \textit{to} field is \code{0x63710c26a9be484581dcac1aacdd95ef628923ab}, which is the attacker-controlled account address. In our system, we monitor the  \textit{to} field of methods \code{eth\_sendTransaction} and \code{eth\_signTransaction} that directly transfer the Ether, and other two methods \code{eth\_estimateGas} and \code{miner\_setEtherBase}. Note that, though the \code{eth\_estimateGas} is used to estimate the gas consumption of a transaction and does not actually send the transaction, executing this method on our honeypot is still considered as a suspicious action since it usually follows a real transaction afterwards. The method \code{miner\_setEtherBase} is used to change the etherbase (or coinbase) account of a miner. This is the address that will be rewarded when a new block is mined by the miner node. By changing this address, the attacker can obtain the reward Ether, on behalf of the miner.

\end{sloppypar}

\begin{listing}[t]
    \usemintedstyle{bw}
    \begin{minted}[frame=single,
        fontsize=\tiny,
        tabsize=2,escapeinside=||]{js}
//The parameters of invoking eth_sendRawTransaction.
{
    "jsonrpc": "2.0", 
    "method": "eth_sendRawTransaction", 
    "params": ["0xf8a682125f8082ea60941a95b271b0535d15fa49932daba31ba6
                12b5294680b844a9059cbb0000000000000000000000000fe07dbd
                07ba4c1075c1db97806ba3c5b113cee00000000000000000000000
                00000000000000000000000000000000000bebc2001ca095e64177
                86f699db2dc195f47662c412bb125b8419b9af030ac237d64c5a92
                50a0357a79a314eecd583f9be2235fd627d85c9af8fe292f9e47d4
                fa261efc0487bc"], 
    "id": 2
}
//The decoded params field of the invocation.
{
    "nonce": 4703, 
    "gasPrice": 0, 
    "gasLimit": 60000, 
    "from":"0x00a329c0648769a73afac7f9381e08fb43dbea72", 
    //This is a smart contract address.
    "to": "0x1a95b271b0535d15fa49932daba31ba612b52946", 
    "value": 0, 
    "data": "0xa9059cbb0000000000000000000000000fe07dbd07ba4c1075c1db9
             7806ba3c5b113cee00000000000000000000000000000000000000000
             00000000000000000bebc200", 
    "v": 28, 
    "r": "0x95e6417786f699db2dc195f47662c412bb125b8419b9af030ac237d64c
          5a9250", 
    "s": "0x357a79a314eecd583f9be2235fd627d85c9af8fe292f9e47d4fa261efc
          0487bc" 
}
    \end{minted}
    \caption{A captured invocation of the method  \code{eth\_sendRawTransaction} and the decoded \textit{params} field of the parameters. } 
    \label{list:attacker-rawtransaction}
\end{listing}

\begin{listing}[t]
    \usemintedstyle{bw}
    \begin{minted}[frame=single,
        fontsize=\tiny,
        tabsize=2,escapeinside=||]{js}
//Function prototype.
Function: transfer(address _to, uint256 _value)
Method ID: 0xa9059cbb   
_to: 0x0fe07dbd07ba4c1075c1db97806ba3c5b113cee0
_value: 200000000(0xbebc200)
    \end{minted}
    \caption{The function to be invoked of the smart contract and its parameters. The \textit{\_to} field contains the attacker's account address that the token will be transferred to.} 
    \label{list:transfer-method}
\end{listing}

Second, we \textit{indirectly} retrieve attackers' account addresses, and use them for grouping. This is the case when attackers steal ERC20 tokens by calling the standard \code{transfer()}~\cite{erc20transfer} function defined in the ERC20 token standard~\cite{erc20tokenstandard}. These addresses are
not in the parameters of the transaction. However, we can obtain them by retrieving parameters of the smart contract method invocation. In the following, we will use a real example to illustrate the steps. Listing~\ref{list:attacker-rawtransaction} shows a captured attack of the RPC request to invoke the method \code{eth\_sendRawTransaction}. We first decode the \textit{params} field to obtain the \textit{to} field (its value is \code{0x1a95b271b0535d15fa49932daba31ba612b52946}). It turns out that this address is not an EOA account, but a smart contract address of the Minereum token~\cite{minereum}, whose peak market capitalization was more than 7 million US dollars in August 2017~\cite{minereumcap}. We further decode the \textit{data} field to retrieve the invoked function in the smart contract and its parameters. The result is shown in Listing~\ref{list:transfer-method}. We can see the malicious account address that receives the stolen ERC20 token is \code{0x0fe07dbd07ba4c1075c1db97806ba3c5b113cee0}.

Third, we group the addresses from the previous two steps based on their IP addresses. We recorded the source IP address of each request, and if multiple attackers' Ethereum account addresses are associated with a same IP address, then we combine these accounts into one group. However, the use of the Tor network may make this strategy ineffective, since their real IP addresses are unknown. In this case, we further group them based on the parameters used in launching the attacks, e.g., the 
\code{id} field in Listing~\ref{list:attacker-address}. For instance, requests from the Tor network have a same \code{id} field. Based on this observation, we believe those requests were initiated from attackers in a same group, i.e., the group $2$ (Table~\ref{tab:grouping}).

Using on previous strategies, we group attackers into $36$ groups. The detailed information of each group is shown in Table~\ref{tab:grouping}. Among them, attackers from $34$ groups (from the group $1$ to the group $34$) were stealing the Ether and other two groups were targeting ERC20 tokens (the group $35$ and $36$).
In the following, we will present the detailed results of attackers' behaviors.

\begin{sloppypar}
    \begin{table*}
        \caption{The result of grouping attackers. }
        \label{tab:grouping}
        \begin{center}
        \scalebox{0.8}{
        \setlength{\tabcolsep}{1mm}{
        \begin{tabular}{lcllllll}
        \toprule
           \#   &  Addresses     &   \# of IP    &   \# of RPC calls  &   Max calls per day &  First capture date & Last capture date &   Days of activity \\
        \midrule
        1   &   \code{0x6a141e661e24c5e13fe651da8fe9b269fec43df0}    &   57  & 72,915,681   &   1,207,185 &   Jul. 1, 2018    &   Feb. 28, 2019   &   178     \\  
            &   \code{0x6e4cc3e76765bdc711cc7b5cbfc5bbfe473b192e}    &       &  &  &  &  &    \\
            &   \code{0x6ef57be1168628a2bd6c5788322a41265084408a}    &       &  &  &  &  &    \\
            &   \code{0x7097f41f1c1847d52407c629d0e0ae0fdd24fd58}    &       &  &  &  &  &    \\
            &   \code{0xe511268ccf5c8104ac8f7d01a6e6eaaa88d84ebb}    &       &  &  &  &  &    \\
\hline  2   &   \code{0x581061c855c24ca63c9296791de0c9a1a5a44fcf}    &   309 &   363,860    &   167,183  &   Jul. 2, 2018    &   Feb. 28, 2019   &   166     \\
            &   \code{0x5fa38ab891956dd35076e9ad5f9858b2e53b3eb5}    &       &  &  &  &  &    \\
            &   \code{0x8cacaf0602b707bd9bb00ceeda0fb34b32f39031}    &       &  &  &  &  &    \\
            &   \code{0xab259c71e4f70422516a8f9953aaba2ca5a585ae}    &       &  &  &  &  &    \\
            &   \code{0xd9ee4d08a86b430544254ff95e32aa6fcc1d3163}    &       &  &  &  &  &    \\
            &   \code{0x88b7d5887b5737eb4d9f15fcd03a2d62335c0670}    &       &  &  &  &  &    \\
            &   \code{0xe412f7324492ead5eacf30dcec2240553bf1326a}    &       &  &  &  &  &    \\
\hline  3   &   \code{0xd6cf5a17625f92cee9c6caa6117e54cbfbceaedf}    &   14  & 13,315,318   &   365,878  &   Jul. 16, 2018   &   Feb. 25, 2019   &   78      \\  
            &   \code{0x21bdc4c2f03e239a59aad7326738d9628378f6af}    &       &  &  &  &  &    \\
            &   \code{0x72b90a784e0a13ba12a9870ff67b68673d73e367}    &       &  &  &  &  &    \\
\hline  4   &   \code{0x04d6cb3ed03f82c68c5b2bc5b40c3f766a4d1241}    &   1   & 101,731,595  &   4,802,304 &   Jul. 1, 2018    &   Nov. 5, 2018    &   63      \\ 
            &   \code{0x63710c26a9be484581dcac1aacdd95ef628923ab}    &       &  &  &  &  &    \\
\hline  5   &   \code{0xb0ec5c6f46124703b92e89b37d650fb9f43b28c2}    &   6   &   326,154    &   9,711    &   Jul. 2, 2018    &   Dec. 3, 2018    &   64      \\  
\hline  6   &   \code{0x1a086b35a5961a28bead158792a3ed4b072f00fe}    &   3   &   6,791,438  &   3,346,904 &   Nov. 1, 2018    &   Feb. 28, 2019   &   21      \\ 
            &   \code{0x73b4c0725c900f0208bf5febb36856abc520de26}    &       &  &  &  &  &    \\
            &   \code{0xaff4778d8d05e9595d540d40607c16f677c73cca}    &       &  &  &  &  &    \\
            &   \code{0xec13837d5e4df793e3e33b296bad8c4653a256cb}    &       &  &  &  &  &    \\
\hline  7   &   \code{0x241946e18b9768cf9c1296119e55461f22b26ada}    &   1   &   7,750,800  &   118,127  &   Jul. 2, 2018    &   Feb. 28, 2019   &   151     \\  
\hline  8   &   \code{0x8652328b96ff12b20de5fdc67b67812e2b64e2a6}    &   2   &   3,569,924  &   281,975  &   Jul. 1, 2018    &   Jul. 30, 2018   &   28      \\  
\hline  9   &   \code{0xff871093e4f1582fb40d7903c722ee422e9026ee}    &   1   &   3,522      &   1,128    &   Jul. 2, 2018    &   Aug. 31, 2018   &   27      \\  
\hline  10  &   \code{0x6230599f54454c695b5cd882064071fc39e6e562}    &   1   &   13         &   13  &   Jul. 5, 2018    &   Jul. 5, 2018    &   1       \\  
\hline  11  &   \code{0x2c5129bdfc6f865e17360c551e1c46815fe21ec8}    &   1   &   618        &   618 &   Jul. 5, 2018    &   Jul. 5, 2018    &   1       \\  
\hline  12  &   \code{0xeb29921d8eb0e32b2e7106afca7f53670e4107e5}    &   1   &   5          &   5   &   Jul. 29, 2018   &   Jul. 29, 2018   &   1       \\  
\hline  13  &   \code{0xe231c73ab919ec2b9aaeb87bb9f0546aa47581b1}    &   1   &   10         &   5   &   Jul. 4, 2018    &   Jul. 17, 2018   &   5       \\  
\hline  14  &   \code{0x5c8404b541881b9999ce89c00970e5e8862f8e88}    &   3   &   80         &   46  &   Jul. 10, 2018   &   Jul. 15, 2018   &   5       \\  
\hline  15  &   \code{0x5e87bab71bbea5f068df9bf531065ce40a86ebe4}    &   1   &   274        &   274 &   Jul. 11, 2018   &   Jul. 11, 2018   &   1       \\  
\hline  16  &   \code{0x97743cc5a168a59a86cf854cf04259abe736006a}    &   3   &   235,213    &   71,521   &   Jul. 10, 2018   &   Jul. 17, 2018   &   8       \\ 
            &   \code{0x9d6d759856bfcabf6f405f308d450b79e16dd4e2}    &       &  &  &  &  &    \\
\hline  17  &   \code{0x02a4347035b7ba02d79238855503313ecb817688}    &   3   &   11,246,017 &   175,417  &   Nov. 13, 2018   &   Feb. 28, 2019   &   97      \\  
            &   \code{0xcb31bea86c3becc1f62652bc8b211fe1bd7f8aed}    &       &  &  &  &  &    \\
\hline  18  &   \code{0xe128bb377f284d2719298b0d652d65455c941b5b}    &   1   &   277        &   147 &   Nov. 12, 2018   &   Nov. 16, 2018   &   4       \\ 
\hline  19  &   \code{0xb744d5f73d27131099efee0b70062de6f770a102}    &   2   &   237,481    &   64,462   &   Dec. 18, 2018   &   Feb. 14, 2019   &   17      \\  
\hline  20  &   \code{0x0e0a930fb51c499b624d6ca56fdd9c95c5bf2e06}    &   2   &   59,842     &   37,608   &   Aug. 4, 2018    &   Feb. 7, 2019    &   38      \\  
            &   \code{0x2c022e9a0368747692b7bd532c435c7a78dc447d}    &       &  &  &  &  &    \\
            &   \code{0x3334f7f8bcf593794b01089b6ff4dc63fe023dfe}    &       &  &  &  &  &    \\
            &   \code{0x884aa595c10b3331ce551c2d9f905e52e21fa0bb}    &       &  &  &  &  &    \\
            &   \code{0xef462edb8880c4fd0738e4d3e9393660b9c5ac72}    &       &  &  &  &  &    \\
\hline  21  &   \code{0x9781d03182264968d430a4f05799725735d9844d}    &   8   &   38,558     &   13,559   &   Aug. 28, 2018   &   Aug. 31, 2018   &   4       \\  
\hline  22  &   \code{0x98c6428fbca6c0ff97570d822dd607f8a55080e5}    &   6   &   270        &   140 &   Aug. 2, 2018    &   Aug. 5, 2018    &   2       \\  
\hline  23  &   \code{0xa0b0209a04398cb61d845148623e68b3eff8f8cb}    &   1   &   135        &   135 &   Jul. 9, 2018    &   Jul. 9, 2018    &   1       \\  
\hline  24  &   \code{0x21d8976138a2b280d441fd7b12456a1193cb2baf}    &   1   &   18,597     &   2,285    &   Aug. 10, 2018   &   Nov. 9, 2018    &   14      \\ 
\hline  25  &   \code{0xfed69981c21b96ff37fc52f9e19849126624ddfd}    &   5   &   963        &   825 &   Aug. 13, 2018   &   Aug. 19, 2018   &   3       \\  
\hline  26  &   \code{0x31c3ecd12abe4f767cb446b7326b90b1efc5bbd9}    &   3   &   440,962    &   49,995   &   Nov. 1, 2018    &   Feb. 14, 2019   &   41      \\ 
\hline  27  &   \code{0x5f622d88cd745ebb8ff2d4d6b707204c65243438}    &   1   &   2,782      &   113 &   Nov. 1, 2018    &   Feb. 28, 2019   &   116     \\  
\hline  28  &   \code{0xf2565682d4ce75fcf3b8e28c002dfc408ab44374}    &   1   &   9          &   3   &   Dec. 22, 2018   &   Jan. 11, 2019   &   5       \\  
\hline  29  &   \code{0xc97663c1156422e2ad33580adab45cad33cf7698}    &   1   &   3,298      &   3,297    &   Feb. 3, 2019    &   Feb. 10, 2019   &   2       \\  
\hline  30  &   \code{0xc6c42a825555fbef74d21b3cb6bfd7074325c348}    &   9   &   73,302     &   4,327    &   Nov. 4, 2018    &   Jan. 6, 2019    &   22      \\  
\hline  31  &   \code{0x454d7320d5751de29074a55ac95bbde312dd7615}    &   1   &   11         &   11  &   Feb. 5, 2019    &   Feb. 5, 2019    &   1       \\  
\hline  32  &   \code{0x4e25e7e76dbd309a1ab2a663e36ac09615fc81eb}    &   1   &   24         &   23  &   Jan. 15, 2019   &   Jan. 16, 2019   &   2       \\  
\hline  33  &   \code{0xa6a21375ca42dcc26237f3e861d58f88fe72eab2}    &   1   &   256        &   256 &   Nov. 27, 2018   &   Nov. 27, 2018   &   1       \\  
\hline  34  &   \code{0xb703ae04fd78ab3b271177143a6db9e00bdf8d49}    &   1   &   1,345      &   72  &   Dec. 30, 2018   &   Feb. 21, 2019   &   32      \\  
\hline  35  &   \code{0x0fe07dbd07ba4c1075c1db97806ba3c5b113cee0}    &   11  &   536,612    &   27,062   &   Jul. 1, 2018    &   Feb. 28, 2019   &   144     \\  
\hline  36  &   \code{0xaa75fb2dcac2e3061a44c831baf0d4c2d4f92fd7}    &   5   &   26,991     &   9,510    &   Jul. 16, 2018   &   Nov. 9, 2018    &   11      \\  
            &   \code{0xffecffe94c3e87987454f2392676ccdb98b926f8}    &       &  &  &  &  &    \\ 
        \bottomrule
        \end{tabular}}}
\end{center}
        \end{table*}

\end{sloppypar}

\begin{table}
    \caption{Most used commands for probing.}
    \begin{center}
    \label{tab:probing}
    \footnotesize
    \begin{tabular}{lll}
    \toprule
    Command& \# of IP addresses &\# of RPC requests\\
    \midrule
    net\_version         & 122 & 4,822,620     \\
    rpc\_modules         & 81  & 3,815       \\
    web3\_clientVersion  & 103  & 4,495,312   \\
    eth\_getBlockByNumber& 325 & 1,190,445    \\
    eth\_blockNumber     & 225 & 27,019,686  \\
    eth\_getBlockByHash  & 214 & 1,633         \\
    \bottomrule
    \end{tabular}
    \end{center}
\end{table}

\subsection{The Analysis of Ether Stealing}
\label{subsec:analysisethstealing}
The attackers from the group $1$ to the group $34$ are stealing the Ether. They are following a three-steps pattern to perform the attack.

\smallskip \noindent
\prompt{Step 1 - Probing potential victims}\tab The first step to launch the attack is to locate potential victims that have insecure HTTP JSON-RPC endpoints. Attackers could obtain potential victims by downloading a list of full Ethereum nodes, or performing a port scanning process to find the machines with target port number ($8545$ in our study) opening. Then attackers issue RPC requests to determine whether the found IP address is an Ethereum node or even a miner node that will be useful to send a zero gas transaction (we will discuss this type of transaction in Section~\ref{subsec:erc20}).

The mostly used commands for Ethereum node probing are shown in Table~\ref{tab:probing}. Specifically, \textit{net\_version} is used to identify the client's current network id to check whether the Ethereum node is running on the mainnet or a testnet. As the name indicates, the usage of the testnet is for testing purpose, and the Ether on this network has no value. By invoking this method, the attacker could find the right targets running the Ethereum mainnet. The \textit{rpc\_modules} command returns all enabled modules.
By probing this information, attackers can get the information of enabled modules and then invoke the APIs inside each module accordingly. Besides the previously discussed two methods, other ones shown in Table~\ref{tab:probing} are also serving the purpose of collecting client information. 

\begin{table}
    \caption{Commands used to prepare attacking parameters.}
    \label{tab:input}
    \begin{center}
    \footnotesize
    \begin{tabular}{lll}
    \toprule
    Command& \# of IP addresses &\# of RPC requests\\
    \midrule
    eth\_accounts            & 615   &   27,040,164  \\
    eth\_coinbase            & 64    &   87,442      \\
    personal\_listAccounts   & 11    &   95          \\
    personal\_listWallets    & 5     &   173,243    \\
    eth\_gasPrice            & 21    &   63,133      \\
    eth\_getBalance          & 493   &   93,585,372   \\
    eth\_getTransactionCount & 63    &   2,411,504   \\
    \bottomrule
    \end{tabular}
    \end{center}
\end{table}

\begin{sloppypar}
\smallskip \noindent
\prompt{Step 2 - Preparing attacking parameters}\tab After locating potential victims, attackers need to prepare the necessary data to launch further attacks. In order to steal the Ether, the attacker needs to send an Ethereum transaction with valid parameters. Specifically, each transaction needs \textit{from\_address} and \textit{to\_address} as the source and destination of a transaction, and other optional ones including \textit{gas}, \textit{gasPrice}, \textit{value} and \textit{nonce}. In order to make the attack succeed, valid parameters should be prepared to steal the Ether.

\begin{figure}[t]
    \centering\includegraphics[width=8cm]{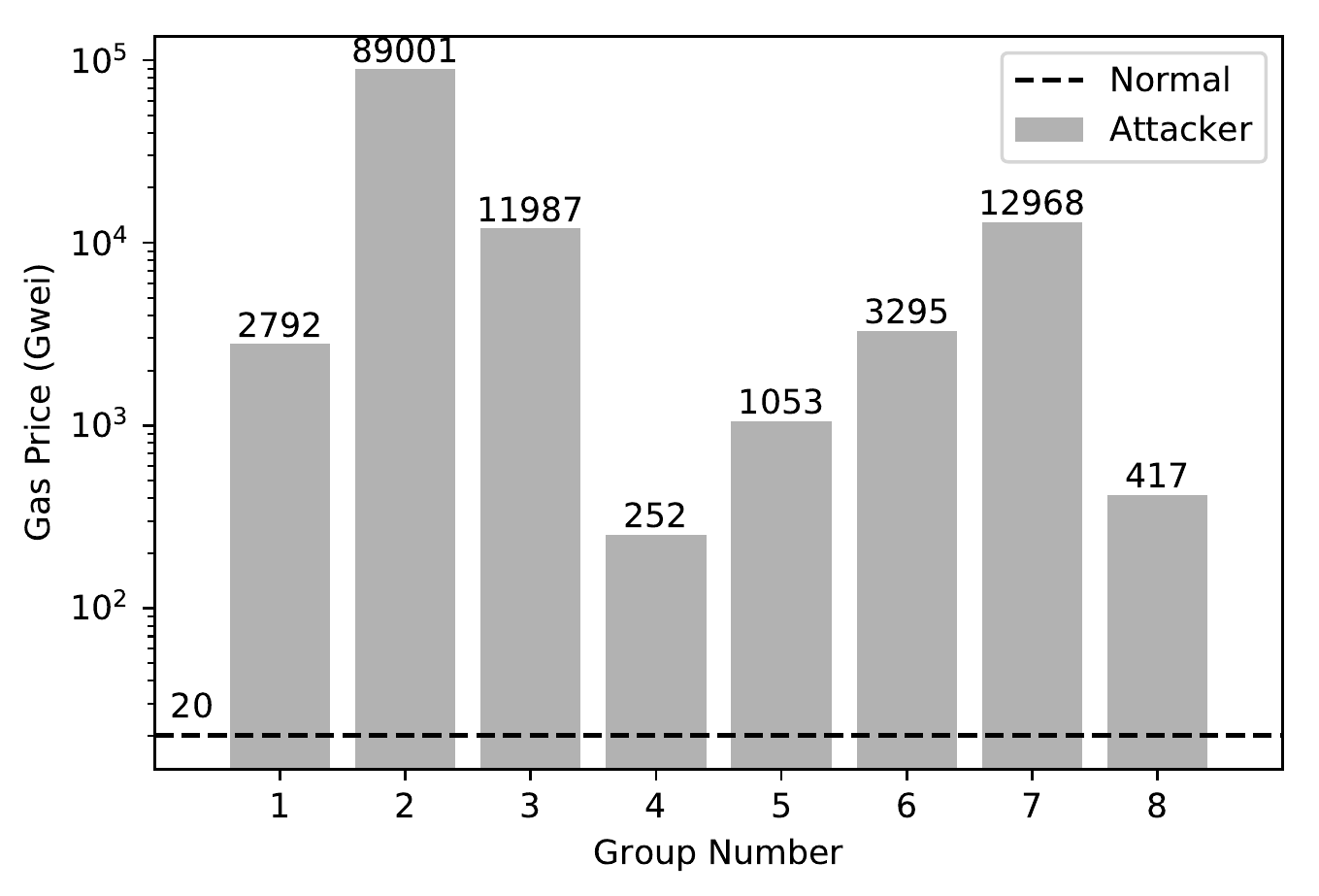}
    \caption{The comparison of the gas price in the transactions of attackers and normal users. The typical gas price is $21$ Gwei, while the gas price of attackers' transactions is much higher.}
    \label{fig:gasprice}
\end{figure}

\begin{itemize}[leftmargin=*]
    \item \textit{from\_address}: The \textit{from\_address} in the transaction is the victim's Ethereum account address. The attacker can obtain this value through invoking the following methods, including \textit{eth\_accounts}, \textit{eth\_coinbase}, \textit{personal\_listAccounts}, \textit{personal\_listWallets}.
    \item \textit{to\_address}: The \textit{to\_address} in the transaction specifies the destination of the transaction. Attackers will set this field to the account under their control.
    \item \textit{value}: This is the value of Ether that will be transferred into the \textit{to\_address}. In order to maximize their income, the attacker tends to transfer all the Ether in the victim's account, leaving a small amount to pay the transaction fee. In order to get the balance of the victim's account, the method \textit{eth\_getBalance} is used.
    \item \textit{gasPrice}: The attacker could set a high \textit{gasPrice} to increase the chance of the transaction being executed (or packed into a block by miners.) For instance, the attacker (\code{0x21bdc4c2f03e239a59aad7326738d9628378f6af}) tends to use a much higher \textit{gasPrice} in the transaction to steal Ether. Figure~\ref{fig:gasprice} shows the gas price of transactions from attackers and normal users. We will illustrate it later in this section. 
\end{itemize}
\end{sloppypar}

\begin{sloppypar}
\smallskip \noindent
\prompt{Step 3 - Stealing Ether}\tab In order to successfully send a transaction, it needs to be signed using the victim's private key. However,
the private key is locked \textit{by default},
and a password is needed to unlock it. We observed two different behaviors that are leveraged by attackers to solve this problem.  

\begin{itemize}[leftmargin=*]
\item {Continuously polling: } Attackers continuously invoke the methods, i.e., \textit{eth\_sendTransaction} or \textit{eth\_signTransaction} in the background. If a legitimate user wants to send a transaction at the same time, then he or she will unlock the account by providing the password. This leaves a small time window that the attacker's attempt to send a transaction will succeed.

However, in order to successfully launch the attack, there are still two challenges. First, the time window is really small. Attackers should happen to invoke the method to send a transaction at the same time when the user is unlocking the account. To increase the chance of a successful attack, the operation to send the transaction should be very frequent. That's the reason of our observation that some attackers are repeatedly invoking the previously mentioned methods at a very high frequency, nearly $50$ requests per-second. Second, since the attacker is sending the transaction at the same time with the user (i.e., when the user is unlocking his account), his transaction may fail if the user's transaction is accepted by the miner at first and the remaining balance of the account will not be sufficient for the attacker's transaction. In order to ensure that his transaction will be accepted by miner in a timely fashion, the attacker will use a much higher value of the \textit{gasPrice} than normal transactions to \textit{bribe} miners.

Figure~\ref{fig:gasprice} shows the gas price of transactions from attackers and normal users. Specifically, we first calculate the average gas price of captured transactions from the group 1 to 8 (the solid line in the Figure). Then we calculate the average gas price of transactions of normal users in six months (the dash line in the Figure). It turns out that the gas price from attacker's transactions is much higher (from $15$ times to $4,500$ times) than the value of a normal transaction. Setting a higher gas price can increase the speed that their transactions are packed into a block. This strategy is very effective, and we have observed several cases that the transaction with a higher \textit{gasPrice} succeed, while the ones with lower \textit{gasPrice} failed~\cite{26b928a,5cdd77e}.

\smallskip 
\item {Brute force cracking: } Besides the polling strategy, some attackers are leveraging the brute force attack to guess the password. Specifically, they try to unlock the account using the password in a predefined dictionary. Since the Ethereum client does not limit the number of wrong password attempts during a certain time period, this attack is effective if the victim uses a weak password. For instance, attackers from the group $11$ leveraged this strategy and tried a dictionary with more than $600$ weak passwords, e.g., \code{qwerty123456}, \code{margarita} and \code{192837465}. Another attacker from the group $1$ took the same way, but only tried one password (\code{ppppGoogle}). The reason why this specific password was used is unknown. However, we think it may be the default password for some customized Ethereum clients.

Interestingly, after a successful try to unlock the account, the attacker will set a relatively long timeout value by invoking the \textit{personal\_unlockAccount}. By doing so, the account will not be locked again in a long time period so that the attacker can perform further attacks much easier. 
\end{itemize}

\end{sloppypar}

\subsection{The Analysis of ERC20 Token Stealing}
\label{subsec:erc20}

\begin{sloppypar}
Attackers from two groups (group $35$ and $36$) are targeting ERC20 tokens. ERC20 is a technical standard used by smart contracts on the Ethereum network to implement exchangeable tokens~\cite{erc20tokenstandard}. The ERC20 token can be viewed as a kind of cryptocurrency that could be sold on some markets, thus becoming valuable targets.

Before illustrating the detailed attacking behaviors, we will first discuss an interesting type of transaction called \code{zero gas transaction}, which we observed in our dataset. It exploits the packing strategy of some miners to send transactions without paying any transaction fee. By using this type of transaction, attackers could perform malicious activities to steal ERC20 tokens from addresses with leaked private key, or exploit the AirDrop mechanism of ERC20 smart contracts to gain extra bonus tokens with nearly \textit{zero} cost.

\end{sloppypar}

\smallskip \noindent
{\prompt{Zero gas transaction}}\tab
Sending a transaction usually consumes gas (Section~\ref{subsec:transaction}). The actual cost is calculated as the product of the amount of gas consumed and the current gas price. The amount of gas consumed during a transaction depends on the instructions executed in the Ethereum virtual machine, while the gas price is specified by the user who sends the transaction. If it is not specified, a default gas price will be used.

\begin{sloppypar}
Interestingly, our honeypot captured many attempts of sending transactions with a zero value in the \code{gasPrice} field. This brings our attention for a further investigation. We want to understand whether such transactions could be successful, and the intentions for sending such transactions. After performing multiple experiments, transactions with a zero gas price received through the p2p network are not accepted by
the miner, and will be discarded as invalid ones. 
However, if such a transaction is created and launched on the
miner node itself (i.e., the node that successfully mines a new block is sending
a zero gas transaction),
then the transaction will be packed into the block by the miner and accepted by the network.


This explains the existences of such transactions captured by our honeypot. In particular, attackers were launching zero gas transactions on every vulnerable Ethereum node, in hope that the node is a miner node that is successfully mining a new block. Though the chance looks really slim, we found several successful cases in reality, e.g., the first several transactions in the block \code{5899499}~\cite{5899499}. Most of the transactions are transferring ERC20 tokens to the address \code{0x0fe07dbd07ba4c1075c1db97806ba3c5b113cee0}, which is a malicious account owned by the attacker in group 20.
\end{sloppypar}

\smallskip \noindent
{\prompt{Attack I: stealing tokens from \textit{fisher} accounts}}\tab The first type of attack is leveraging the zero gas transactions to steal tokens from \textit{fisher} accounts. In order to understand this attack, we first explain what the \textit{fisher} account is in the following.

\begin{sloppypar}
The \textit{fisher} account means that some attackers intentionally leak the private key of their Ethereum accounts on Internet. They also transfer some ERC20 tokens to the accounts as the bait. Since the private key of the account is leaked, other users could use the private key to transfer out the ERC20 tokens. However, there is one problem in this process. In order to transfer the ERC20 tokens, the account should have some Ethers to pay the transaction fee. As a result, one may transfer some Ethers into this account, in hope to get the ERC20 tokens. Unfortunately, after transferring the Ether into this account, the Ether will be transferred out to some accounts immediately by attackers. That's the reason why such an account is called the \textit{fisher} account. The main purpose of leaking the private key is to seduce others transferring Ether into the \textit{fisher} account. 

For instance, there is a \textit{fisher} account whose address is \code{0xa8015df1f65e1f53d491dc1ed35013031ad25034}~\cite{A8015DF1}. The attacker bought $75,000$ ICX (a ERC20 token) as the fishing bait that values around $66,000$ US dollars. \textit{Occasionally}, the fisher released the private key of that account on the Internet. Anyone who transfers the Ether into this account and hopes to obtain the ICX token will be trapped to lose the transferred Ether.

Interestingly, by leveraging the zero gas transaction previously discussed, attackers could steal the ERC20 tokens in the  \textit{fisher} account. Specifically, attackers could send the transactions to transfer the ERC20 tokens in the \textit{fisher} account with zero gas price. If the transaction is successful, then the attackers will obtain the ERC20 tokens without any cost.

In our dataset, the user in group $35$ (the address is \code{0x0fe07dbd07ba4c1075c1db97806ba3c5b113cee0}) was performing this type of attack. In total, the attacker sent $61,158$ RPC requests, stealing $161$ different types of ERC20 tokens. 
We show the detailed information of the top ten ERC20 tokens that this attacker is targeting in Table~\ref{tab:erc20}. We observed several different IP addresses (62.75.138.194, 77.180.167.78, 77.180.200.1, 92.231.160.88, 92.231.169.137, 95.216.158.152 and etc.) from this attacker.  



 \begin{table}
        \caption{The top ten ERC20 tokens that attackers are targeting.}
        \label{tab:erc20}
        \begin{center}
        \scalebox{0.8}{
        \setlength{\tabcolsep}{1mm}{
        \begin{tabular}{cp{35px}p{35px}}
        \toprule
	 \multirow{2}*{ERC20 token addresses} 	 & \# of RPC requests & Token name\\
    \midrule
    \code{0x1a95b271b0535d15fa49932daba31ba612b52946}  &  11,788 &  MNE  \\
    \code{0xee2131b349738090e92991d55f6d09ce17930b92}  &  8,998  &  DYLC  \\
    \code{0x0775c81a273b355e6a5b76e240bf708701f00279}  &  8,099  &  BUL  \\
    \code{0xbdeb4b83251fb146687fa19d1c660f99411eefe3}  &  7,735  &  SVD  \\
    \code{0x0675daa94725a528b05a3a88635c03ea964bfa7e}  &  7,359  &  TKLN  \\
    \code{0x87c9ea70f72ad55a12bc6155a30e047cf2acd798}  &  7,058  &  LEN  \\
    \code{0x4c9d5672ae33522240532206ab45508116daf263}  &  5,510  &  VGS  \\
    \code{0x23352036e911a22cfc692b5e2e196692658aded9}  &  4,011  &  FDZ  \\
    \code{0xc56b13ebbcffa67cfb7979b900b736b3fb480d78}  &  2,219  &  SAT  \\
    \code{0x89700d6cd7b77d1f52c29ca776a1eae313320fc5}  &  1,708  &  PMD  \\
        \bottomrule
        \end{tabular}}}
\end{center}
\vspace{-0.4em}
        \end{table}
\end{sloppypar}

\smallskip \noindent
{\prompt{Attack II: Exploiting the airdrop mechanism}\tab Airdrop is a marketing strategy that the token holders would receive bonus tokens based on some criteria, e.g., the amount of total tokens they hold. The conditions to send out bonus tokens depend on the individual token maintainer. 

\begin{sloppypar}
Some attackers are leveraging the zero gas transaction to obtain the free LEN tokens. Specifically, the LEN token has an airdrop strategy that if a new user A sends any amount of LEN token to the user B, then both A and B will be rewarded with $18,895$ LEN tokens. Hence, the attacker could create a large number of new accounts, and then transfer LEN tokens to the attacker's address (address \code{0xffecffe94c3e87987454f2392676ccdb98b926f8} in group $36$). By doing so, the new account will receive a bonus token, which will be transferred to the attacker's account, while at the same time the attacker's account will also receive the bonus. We observed many attempts of such transactions using zero gas price, with $7,058$ different source account addresses and one destination address (the attacker's address). This transaction does not consume any gas, and the attacker could be rewarded with ERC20 tokens. By using this method, the attacker even becomes a large holder of this token ($2.4\%$)~\cite{ffecffe94}.
\end{sloppypar}
}
\section{Transaction Analysis}
\label{sec:profit}

\begin{table*}[!p]
    \caption{Our estimation of attackers' profits in Ether and US dollars. The price of one Ether is around $139$ US dollars (March, 2019). We remove the addresses with zero profit from the table.}
    \label{tab:gains}
	\begin{center}
        \scalebox{1}{
    \begin{tabular}{c|c|c|c|c|c|}
    \toprule
 
    \multirow{2}*{\#} & \multirow{2}*{Addresses}  & \multicolumn{2}{c|}{Malicious} & \multicolumn{2}{c|}{Plus Suspicious}  \\
    \cline{3-6}
    ~ &~ & Ether & USD & Ether & USD  \\
\midrule
1	&\code{0x6a141e661e24c5e13fe651da8fe9b269fec43df0}	&	116.91&\$16,280.23&814.45&\$113,412.00\\
	&\code{0x6e4cc3e76765bdc711cc7b5cbfc5bbfe473b192e}	&	56.16&\$7,820.34&794.67&\$110,657.59\\
	&\code{0x6ef57be1168628a2bd6c5788322a41265084408a}	&	37.79&\$5,261.74&1,420.06&\$197,743.19\\
	&\code{0x7097f41f1c1847d52407c629d0e0ae0fdd24fd58}	&	281.44&\$39,191.07&1,331.74&\$185,444.98\\
	&\code{0xe511268ccf5c8104ac8f7d01a6e6eaaa88d84ebb}	&	152.26&\$21,201.86&1,332.53&\$185,554.52\\
	&\code{0x8652328b96ff12b20de5fdc67b67812e2b64e2a6}	&	37.75&\$5,256.18&1,066.31&\$148,483.43\\
	&\code{0xff871093e4f1582fb40d7903c722ee422e9026ee}	&	0.00&\$0.69&9.34&\$1,300.01\\
2	&\code{0x5fa38ab891956dd35076e9ad5f9858b2e53b3eb5}	&	48.24&\$6,716.88&94.28&\$13,129.14\\
	&\code{0x8cacaf0602b707bd9bb00ceeda0fb34b32f39031}	&	0.00&\$0.14&10.66&\$1,483.80\\
	&\code{0xab259c71e4f70422516a8f9953aaba2ca5a585ae}	&	2.53&\$351.64&4.18&\$581.92\\
	&\code{0xd9ee4d08a86b430544254ff95e32aa6fcc1d3163}	&	54.12&\$7,535.80&55.72&\$7,759.26\\
	&\code{0x88b7d5887b5737eb4d9f15fcd03a2d62335c0670}	&	0.24&\$33.41&0.24&\$33.41\\
	&\code{0xe412f7324492ead5eacf30dcec2240553bf1326a}	&	0.24&\$33.96&0.24&\$33.96\\
	&\code{0x241946e18b9768cf9c1296119e55461f22b26ada}	&	1.53&\$213.74&1.53&\$213.74\\
	&\code{0x9781d03182264968d430a4f05799725735d9844d}	&	50.32&\$7,006.89&61.47&\$8,560.18\\
4	&\code{0x04d6cb3ed03f82c68c5b2bc5b40c3f766a4d1241}	&	2.38&\$331.13&2.38&\$331.13\\
	&\code{0x63710c26a9be484581dcac1aacdd95ef628923ab}	&	19.44&\$2,706.79&38.88&\$5,413.47\\
	&\code{0xb0ec5c6f46124703b92e89b37d650fb9f43b28c2}	&	0.87&\$120.84&1.64&\$227.89\\
6	&\code{0x1a086b35a5961a28bead158792a3ed4b072f00fe}	&	80.22&\$11,170.51&4,821.68&\$671,419.10\\
	&\code{0x73b4c0725c900f0208bf5febb36856abc520de26}	&	1.10&\$153.12&1.10&\$153.12\\
	&\code{0xec13837d5e4df793e3e33b296bad8c4653a256cb}	&	1.62&\$226.21&1.62&\$226.21\\
11	&\code{0x2c5129bdfc6f865e17360c551e1c46815fe21ec8}	&	113.93&\$15,864.25&506.86&\$70,580.16\\
15	&\code{0x5e87bab71bbea5f068df9bf531065ce40a86ebe4}	&	0.05&\$6.42&0.05&\$6.42\\
17	&\code{0x02a4347035b7ba02d79238855503313ecb817688}	&	4.30&\$598.46&4.30&\$598.46\\
	&\code{0xcb31bea86c3becc1f62652bc8b211fe1bd7f8aed}	&	0.21&\$29.19&0.21&\$29.19\\
	&\code{0xd6cf5a17625f92cee9c6caa6117e54cbfbceaedf}	&	2,030.19&\$282,704.44&2,030.19&\$282,704.44\\
	&\code{0x21bdc4c2f03e239a59aad7326738d9628378f6af}	&	357.78&\$49,820.26&58,692.91&\$8,172,988.20\\
	&\code{0x72b90a784e0a13ba12a9870ff67b68673d73e367}	&	558.32&\$77,746.63&59,298.45&\$8,257,309.40\\
26	&\code{0x31c3ecd12abe4f767cb446b7326b90b1efc5bbd9}	&	0.10&\$13.23&0.10&\$13.23\\
28	&\code{0xf2565682d4ce75fcf3b8e28c002dfc408ab44374}	&	173.99&\$24,228.78&866.10&\$120,604.60\\
	&\code{0xb703ae04fd78ab3b271177143a6db9e00bdf8d49}	&	8.02&\$1,116.77&8.02&\$1,116.77\\
30	&\code{0xc6c42a825555fbef74d21b3cb6bfd7074325c348}	&	1.50&\$208.36&1.50&\$208.36\\
32	&\code{0x4e25e7e76dbd309a1ab2a663e36ac09615fc81eb}	&	0.04&\$6.27&0.05&\$7.34\\
\midrule
Total&&4,193.58&\$583,956.23&133,273.46&\$18,558,328.61\\

\bottomrule
\end{tabular}}
\end{center}
\end{table*}

After capturing malicious accounts and analyzing the detailed attackers' behaviors, we further estimate profits of attackers. Though we can directly get the estimation by calculating the income of malicious accounts, attackers may use other account addresses that have not been captured by our honeypot. We call these addresses that are potentially controlled by attackers as suspicious accounts.





In our system, we take the following steps to detect suspicious accounts. The basic idea is if the attacker transfers the Ether from a malicious account to any other account, it is highly possible that the destination account is connected with the attacker. The attacker has no reason to transfer the Ether to an account that has no relationship with. Note that, the attacker could transfer the Ether to a cryptocurrency market, where he can exchange it with other types of cryptocurrencies or real currency. These markets should be removed from suspicious accounts in our study~\footnote{We obtained the addresses of cryptocurrency markets from the Etherscan website~\cite{Etherscan}.}. 

\begin{figure}[t]
    
    \includegraphics[scale=0.3]{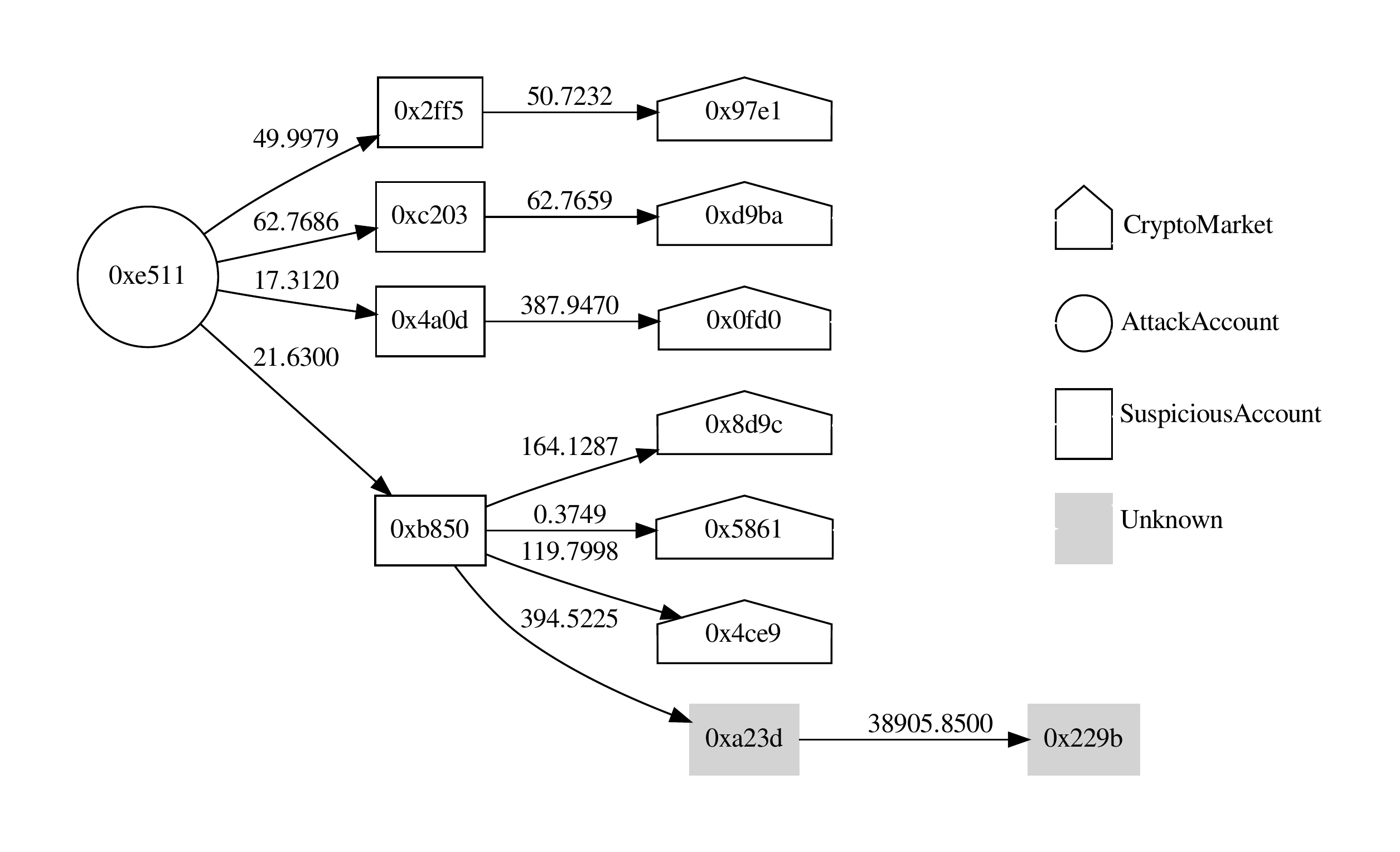}
    \caption{One example of detecting suspicious accounts through the transaction analysis. The house ones are the cryptocurrency markets, the circle one is the malicious account. The box ones without background color are suspicious accounts, while the ones with gray background are unknown accounts.}
    \label{fig:e511}
\end{figure}

\begin{sloppypar}

To this end, we used a similar idea of the taint analysis~\cite{schwartz:2010:dynamic} to find suspicious accounts. Specifically, we treat malicious accounts captured by our system as the taint sources, and propagate the taint tags through the transaction flows until reaching the taint sinks, i.e., the cryptocurrency markets. We also stop this process if the number of accounts traversed reaches a certain threshold. In our study, we use $3$ as the threshold. All the accounts in the path from the taint source to the taint sink are considered tainted and suspicious, as long as the endpoint is a cryptocurrency market. Other nodes are marked as unknown ones, since we do not have further knowledge about whether the nodes are suspicious or not. 
Figure~\ref{fig:e511} shows an example of this process to detect the suspicious accounts from the malicious one \code{0xe511268ccf5c8104ac8f7d01a6e6eaaa88d84ebb}. In this figure, the cryptocurrency market nodes are marked in the house symbol, and the original attacker we captured is marked as a circle. Box nodes without background color are the suspicious accounts we identified, and others with gray background color are unknown addresses. The line between two nodes denotes transactions between them. We also put the number of Ether transferred above the line. In total, we identified $113$ suspicious addresses, and $936$ unknown ones, respectively.
\end{sloppypar}

After that, we estimate attackers' profits. We first calculate the lower bound of profits by only considering the income of the malicious accounts.
Since our honeypot observed their behaviors of stealing the Ether, we have a high confidence that these malicious accounts belong to attackers. Then we add the income of suspicious addresses into consideration. Since these addresses are not directly captured by our honeypot, we do not have a hard evidence that they belong to attackers. However, they may be connected with or controlled by attackers. Table~\ref{tab:gains} shows the estimated  profits. We remove the addresses with zero profit from the table (e.g., addresses in the group 10), and we do not count the attackers from the group $35$ and $36$ since they are targeting ERC20 tokens, whose value are hard to estimate due to the dramatic change of the token price. It's worth noting that, the actual income of attackers are far more than the value shown in the table since there are many attacks in the wild that were not captured by our system.
\section{Discussion}
\label{sec:discussion}

Though we have adopted several ways to make our honeypot an interactive one, cautious attackers can still detect the existence of our honeypot and do not perform malicious activities thereafter. For instance, the attacker can first send a small amount of Ether to a newly generated address and then observe the return value (the transaction hash) of this transaction. Since the transaction to send the Ether in our honeypot does not actually happen, the return value is an invalid one (a randomly generated value). The attackers can also simply send some uncommon commands and observe the return value to detect the honeypot. Nevertheless, it is
an open research question to propose more effective countermeasures to improve the honeypot.


In this paper, we take a conservative way to detect suspicious accounts and estimate profits of attackers. 
Specifically, we leverage the knowledge of whether an address belongs to a cryptocurrency market and mark the tainted accounts whose Ether eventually flows into cryptocurrency markets as suspicious. However, the knowledge of the mapping between addresses and cryptocurrency markets is incomplete, since these addresses are manually labelled. Some suspicious accounts may be identified as unknown ones, hence introducing false negatives to our work. Moreover, our estimation is based on the attacker's addresses collected by our honeypot (in six months). There do exist attackers missed by our system, and profits of these attackers are not included in our estimation. We believe the total income of attackers in the wild is much higher than our estimation.

In this paper, attackers are exploiting the unprotected JSON-RPC interface to launch attacks. Though it is simple to fix the problem by changing the configuration of the Ethereum client, we are surprised by the fact that
$7\%$ of Ethereum nodes are still vulnerable. Specifically, 
we performed a port scanning to the $15,560$ Ethereum public nodes~\cite{Ethernodes} and found that around $1,000$ of them are reachable through
the RPC port without any authentication~\footnote{For ethical reasons, we did not perform any RPC calls that may cause damages to those nodes.}. This fact demonstrated the severity of this problem, and advocates the need to
have a better understanding of this issue in the community (the purpose
of our work.)

\section{Related work}
\label{sec:related}

\smallskip \noindent
{\prompt{Honeypot}}\tab
Honeypot systems have been widely used to capture and understand attacks by capturing malicious activities~\cite{Nepenthes, Collapsar, honeyd, Honeypotfarm}. For instance, HoneyD~\cite{honeyd} is one of the best-known honeypot projects. It can simulate the network stack of many operating systems and arbitrary routine topologies, thus making it a highly interactive one. Collapsar can manage a large number of interactive honeypots, e.g., Honeypot farms. The concept of honeypot has been adopted to detect attacks to IoT devices~\cite{IoTPOT, IoTCandyJar, SIPHON, honeypotcamera} and mobile devices~\cite{mobilehoneypot}. Our system is working towards Ethereum clients, which have different targets with previous systems. The general idea of attracting attacks and logging behaviors are similar, but with different challenges.

\smallskip \noindent
{\prompt{Security issues of smart contracts}}\tab
One reason that Ethereum is becoming popular is its support for smart contracts. Developers can use contracts to develop
decentralized apps (or Dapps), including the lottery game,
or digital tokens. Since its introduction, security issues
of smart contracts have been widely studied by previous researchers. 
Atzei et al. systematically analyzed the security vulnerabilities of Ethereum smart contracts, and proposed several common pitfalls when programming the smart contracts~\cite{atzei2017survey}. For instance, the stack size of the Ethereum virtual machine is limited, attackers could leverage this to hijack the control flow of the smart contracts. A system called teEther~\cite{TEETHER} is proposed to automatically generate the exploits to attack the vulnerable smart contracts. The evaluation showed that among $38,757$ unique smart contracts, $815$ of them could be automatically exploited.

To mitigate the threats, some tools to analyze the smart contracts~\cite{kalra2018zeus, zhou2018erays, Securify, Mythril, MAIAN, DataEther} or fix the
vulnerable contracts~\cite{Sereum} have been proposed. For instance, Oyente~\cite{Oyente} is a system that can automatically detect smart contracts vulnerabilities using the symbolic execution engine. Moreover, it can make the smart contracts less vulnerable by proposing some new semantics to the Ethereum virtual machine. 
Sereum~\cite{Sereum} automatically fixes the reentrancy vulnerabilities in
smart contracts by modifying the Ethereum virtual machine.
Erays~\cite{zhou2018erays} is a tool to analyze the smart contracts without the requirement of the source code. In particular, it translates the bytecode to the high-level code that is readable for manual analysis. Securify~\cite{Securify} automatically proves smart contract behaviors as safe or unsafe. Other similar tools to analyze smart contracts include Mythril~\cite{Mythril} and Maian~\cite{MAIAN}.

\section{Conclusion}
\label{sec:conclusion}

In this paper, we performed a systematic study to understand the cryptocurrency stealing on Ethereum. To this end, we first designed and implemented a system that captured \textit{real} attacks, and further analyzed the attackers' behaviors and estimated their profits. We report our findings in the paper and release the dataset of attacks (https://github.com/zjuicsr/eth-honey) to engage the whole research community.

\smallskip \noindent
\prompt{Acknowledgements}\tab
The authors would like to thank the anonymous
reviewers for their insightful comments that
helped improve the presentation of this paper.
Special thanks go to Siwei Wu, Quanrun Meng for their constructive suggestions and feedbacks.
This work was partially supported by Zhejiang Key R\&D Plan (Grant No. 2019C03133), the National Natural
Science Foundation of China under Grant 61872438,
the Fundamental Research Funds
for the Central Universities,
the Key R\&D Program of Shanxi Province of China under
Grant 2019ZDLGY12-06.
Any opinions, findings,
and conclusions or recommendations expressed in this material are those of the authors and do
not necessarily reflect the views of funding agencies.

{\footnotesize \bibliographystyle{plain}
\bibliography{main}}

\end{document}